\documentclass[a4paper,UKenglish,cleveref, autoref, thm-restate]{lipics-v2019}

\usepackage{graphicx}
\usepackage{amsfonts}
\usepackage{amsmath}
\usepackage{xcolor}
\usepackage{multicol}

\title{The Sitting Closer to Friends than Enemies Problem in Trees}

\titlerunning{The SCFE Problem in Trees}

\author{Rosa Becerra}{Departamento de Ingenier\'ia Matem\'atica, Facultad de Ciencias F\'isicas y Matem\'aticas, Universidad de Concepci\'on, Chile.}{rbecerra@udec.cl}{}{}

\author{Christopher Thraves Caro}{Departamento de Ingenier\'ia Matem\'atica, Facultad de Ciencias F\'isicas y Matem\'aticas, Universidad de Concepci\'on, Chile.}{cthraves@ing-mat.udec.cl}{https://orcid.org/0000-0002-9909-5315}{}

\authorrunning{R. Becerra and C. Thraves Caro} 

\Copyright{Rosa Becerra and Christopher Thraves Caro} 



\category{} 

\relatedversion{} 

\supplement{}

\acknowledgements{\emph{Rosa Becerra} presented some of these results in her thesis to obtain the degree of Mathematical Engineer given by the 
Departamento de Ingenier\'ia Matem\'atica, Facultad de Ciencias F\'isicas y Matem\'aticas, Universidad de Concepci\'on}

\nolinenumbers 

\hideLIPIcs  

\EventEditors{John Q. Open and Joan R. Access}
\EventNoEds{2}
\EventLongTitle{42nd Conference on Very Important Topics (CVIT 2016)}
\EventShortTitle{CVIT 2016}
\EventAcronym{CVIT}
\EventYear{2016}
\EventDate{December 24--27, 2016}
\EventLocation{Little Whinging, United Kingdom}
\EventLogo{}
\SeriesVolume{42}
\ArticleNo{23}
\begin{document}

\maketitle
\begin{abstract}
A metric space $\mathcal{T}$ is a \emph{real tree} if for any pair of points $x, y \in \mathcal{T}$ all topological embeddings $\sigma$  of the segment $[0,1]$ into $\mathcal{T}$, such that $\sigma (0)=x$ and $\sigma (1)=y$, have the same image (which is then a geodesic segment from  $x$ to $y$). A \emph{signed graph} is a graph where each edge has a positive or negative sign. The \emph{Sitting Closer to Friends than Enemies} problem in trees has a signed graph $S$ as an input.  The purpose is to determine if there exists an injective mapping (called \emph{valid distance drawing}) from $V(S)$ to the points of a real tree such that, for every $u \in V(S)$, for every positive neighbor $v$ of $u$, and negative neighbor $w$ of $u$, the distance between $v$ and $u$ is smaller than the distance between $w$ and $u$.

 In this work, we show that a complete signed graph has a valid distance drawing in a real tree if and only if its subgraph composed of all (and only) its positive edges has an intersection representation by unit balls in a real tree. Besides, as an instrumental result, we show that a graph has an intersection representation by unit balls in a real tree if and only if it has an intersection representation by proper balls, and if and only if it has an intersection representation by arbitrary balls in a real tree.  
\end{abstract}

\section{Introduction}\label{sec:intro}
 A \emph{signed graph} is an undirected graph with a positive or negative sign associated with its edges. Signed graphs have attracted increasing attention due to their capability of representing many real-world relations \cite{leskovec2010signed,tang2016survey}. Embedding signed graphs in low-dimensional metric spaces is of particular interest due to applications in clustering, link prediction, and network visualization \cite{kunegis2010spectral,wang2017signed,zheng2015spectral}.

The \emph{Sitting Closer to Friends than Enemies} (SCFE) problem aims to find a meaningful representation of a signed graph that reveals patterns among the data it represents. Formally, the SCFE problem is to find an injection in a metric space of the vertex set of a signed graph such that for every pair of incident edges with different signs, the end vertices of the positive edge are injected closer than the end vertices of the negative edge, in the metric of the space. Such an injection is called a \emph{valid distance} drawing.  Spaen et al. in \cite{Spaen2019} proved that every signed graph on $n$ vertices has a valid distance drawing in $\mathbb{R}^{n-2}$. Nevertheless, and unfortunately for visualization purposes, they also proved the existence of a signed graph on $n$ vertices without a valid distance drawing in $\mathbb{R}^k$ for $k<\lfloor\log_5(n-3)\rfloor+1$. Therefore, the number of dimensions required to represent any signed graph on $n$ vertices is too large if we want to plot this representation in a two-dimensional \emph{paper}. 

In this document, we continue with the quest to find a meaningful representation for signed graphs in low-dimensional metric spaces. In this case, we consider metric spaces with a tree-like structure. In other words, the metric spaces in consideration are the union of simple open curves such that for every two points in the space, there is a unique shortest path between them, and the distance is the length of that shortest path. Such metric spaces are known as \emph{real trees}.

We show that a complete signed graph has a valid distance drawing in a real tree if, and only if, its positive subgraph has an intersection representation by  balls in a real tree (also known as NSI graphs), where the positive subgraph of a signed graph is the subgraph composed of all, and only, its positive edges. As an instrumental result, we show that the following three families of graphs are equivalent: the family of graphs with an intersection model of unit balls on a real tree, the family of graphs with an intersection model of proper balls on a real tree, and the family of graphs with an intersection model of arbitrary balls on a real tree. 

We organize the rest of the document as follows. In Section \ref{sec:def}, we present definitions and notations; we also state the main results of this document. In Section \ref{sec:relwork}, we present the previous results of the SCFE problem as they relate to our contributions.  In Section \ref{sec:4statements}, we show the equivalence between the three families of intersection graphs on real trees. This characterization is valuable on its own, but, in our case, it is also instrumental. We use this characterization to prove the result presented in Section \ref{sec:complete}, which characterizes complete signed graphs with a valid distance drawing in a real tree. We close this document in Section \ref{sec:conclusion} with concluding remarks and a discussion regarding open problems. 

\section{Definitions}\label{sec:def}
We use $G=(V,E)$ to denote a simple undirected graph, where $V$ is the vertex set of $G$ and $E$ is its edge set. Given two vertices $u$ and $v$ in $V$, we use $\{u,v\}$ to denote the edge between vertices $u$ and $v$. If $\{u,v\} \in E$, we say that $u$ and $v$ are \emph{neighbors}, or \emph{adjacent}. 

A \emph{signed graph} is a graph such that its edge set is partitioned in two subsets, the set of positive edges, and the set of negative edges. We use $S=(V, E^+\cup E^-)$ to denote a signed graph, where $\emptyset = E^+\cap E^-$. Given two vertices $u$ and $v$ in $V$, we say that they are \emph{friends} or \emph{positive neighbors} if $\{u,v\}\in E^+$. Equivalently, we sat that $u$ and $v$ are \emph{enemies} or \emph{negative neighbors} if $\{u,v\} \in E^-$. The \emph{positive neighborhood} of a vertex $u\in V$ is the set $N^+(u):=\{v \in V: \{u,v\}\in E^+\}$, and its \emph{closed positive neighborhood} is the set  $N^+[u]:=N^+(u)\cup \{u\}$.
 
The \emph{positive subgraph} of a signed graph $S=(V,E^+\cup E^-)$ is the graph $S^+:=(V,E^+)$ on the same set of vertices and containing all, and only, its positive edges. Note that $S^+$ can be seen as a graph, not necessarily signed, since all its edges have the same sign (positive). 

\begin{figure}[t]
\centering
\includegraphics[width=0.4\textwidth]{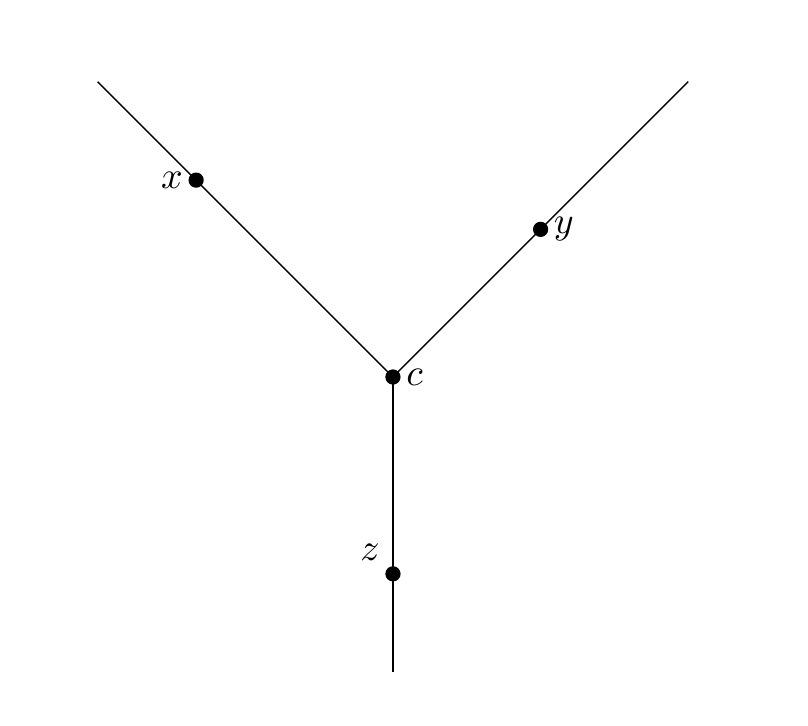}
\caption{This figure illustrates the main characteristic of a real tree. 
For every three points $x,y$ and $z$ there exists a point $c$ such that the 
shortest path between $x$ and $z$ and the shortest path between $y$ and $z$ 
intersect in the segment between $c$ and $z$, and $c$ belongs to the shortest path between $x$ and $y$.
Such point $c$ is called the \emph{center} of the points $x$, $y$ and $z$.}
\label{fig:space}
\end{figure}

In this document, we are interested in metric spaces with a tree-like structure. Let $(\mathcal{T},d)$ be a metric space. A \emph{path} in $\mathcal{T}$ is a continuous function $\sigma$ from the unit interval $[0,1]$ to $\mathcal{T}$. The \emph{extreme points} of a path $\sigma$ are the points $\sigma(0)$ and $\sigma(1)$ in  $\mathcal{T}$. We say that a path $\sigma$ \emph{connects} points $x$ and $y$ in $\mathcal{T}$ if $x$ and $y$ are the extreme points of $\sigma$. The length of a path $\sigma$ in $\mathcal{T}$ is the length along the curve defined by  the image of $\sigma$. We say that $(\mathcal{T},d)$ is \emph{path-connected} if there exists a path connecting any two points in $\mathcal{T}$. A metric space $(\mathcal{T},d)$ is a \emph{real tree} (or {\it $\mathbb{R}$-tree}) if $\mathcal{T}$ is the union of open, simple curves such that it is path-connected and for every three points $x,y$ and $z$ in $\mathcal{T}$, there exists a point $c$ such that the shortest path between $z$ and $x$, and the shortest path between $z$ and $y$ intersect in the shortest path between $z$ and $c$, and $c$ belongs to the shortest path between  $x$ and $y$. We call such point $c$ the \emph{center} of the points $x$, $y$ and $z$. An illustration of this situation is shown in Figure \ref{fig:space}.

\begin{definition}
Let $(\mathcal{T}, d)$ be a metric space, and $S=(V,E^+\cup E^-)$ be a signed graph.
We say that an injection  $D:V\rightarrow \mathcal{T}$ of the vertex set of $S$ into $\mathcal{T}$ is a \emph{valid distance drawing} of $S$ if, for every pair of incident edges with different signs $\{u,v\} \in E^+$ and $\{u,w\} \in E^-$,
\begin{equation}\label{eq:vdd}
d(D(u),D(v)) < d(D(u),D(w)).
\end{equation}
\end{definition}

\begin{definition}
The \emph{SCFE problem in trees} is to decide if a given signed graph $S$ has a valid distance drawing in a real tree $\mathcal{T}$, and, in case of existence, find such $\mathcal{T}$ and such valid distance drawing. 
\end{definition}
 
 Throughout this document, we also use some families of intersection graphs, i.~e., graphs that represent the pattern of intersections of a family of sets. In fact, given a family of sets $\mathcal{B}=\{B_1,B_2,\ldots B_n\}$, the \emph{intersection graph} $G(\mathcal{B})=(V,E)$  of $\mathcal{B}$ is the graph where $V=\{1,2,\ldots,n\}$ has one vertex $u$ per each set $B_u$ in $\mathcal{B}$, and $\{u,v\} \in E$ if $B_u \cap B_v \neq \emptyset$. Classic examples of intersection graphs are \emph{interval graphs}, the intersection graphs of intervals on the real line,  \emph{unit interval} graphs, the intersection graphs of intervals of the same length on the real line, and \emph{proper interval} graphs, the intersection graphs of intervals on the real line such that no interval fully contains another interval. If $G$ is the intersection graph of a family of sets $\mathcal{B}$, we say that $\mathcal{B}$ is an \emph{intersection model} for $G$. 
  
  The \emph{ball} centered at $x$ with radius $r$ on a tree $(\mathcal{T},d)$ is the set $B(x,r):=\{y \in \mathcal{T}:d(x,y)\leq r\}$ of points in $\mathcal{T}$. A point $y$ in $B(x,r)$ such that $d(x,y) = r$ is called an \emph{extreme point} of $B(x,r)$. A graph is a \emph{Neighborhood Subtree Intersection} (NSI) graph if it is the intersection graph of a family of balls on a real tree. A graph is a \emph{unit-NSI} graph if it is the intersection graph of a family of balls, all with the same radius, on a real tree. A graph is a \emph{proper-NSI} graph if it is the intersection graph of a family of balls on a real tree such that no ball is fully contained in another. Their respective intersection models are called \emph{NSI 
 model}, \emph{unit-NSI model}, and \emph{proper-NSI model}, respectively.
 
\begin{figure}[t]
  \centering
  \includegraphics[width=0.6\textwidth]{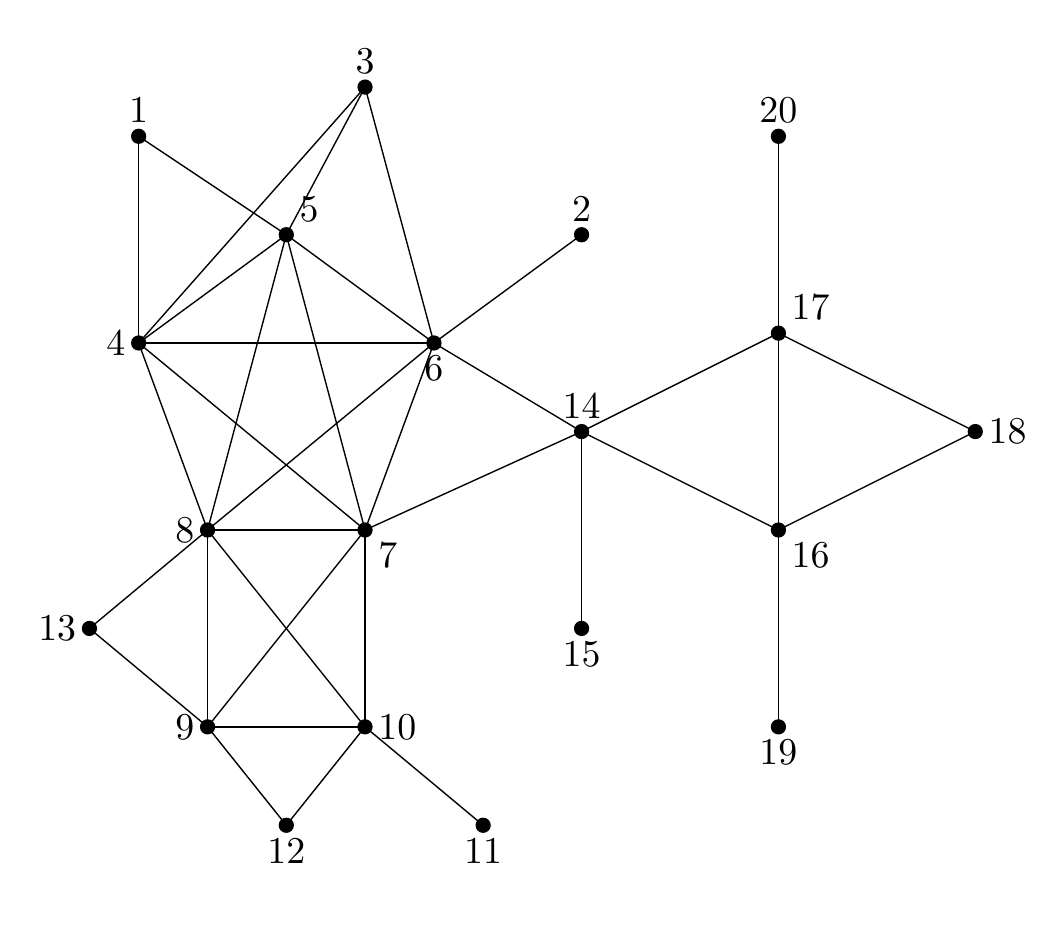}
 \caption{In this figure we show an example of an NSI graph.}
\label{fig:stronglychordal}
  \end{figure}

\begin{figure}[t]
\centering
\includegraphics[width=\textwidth]{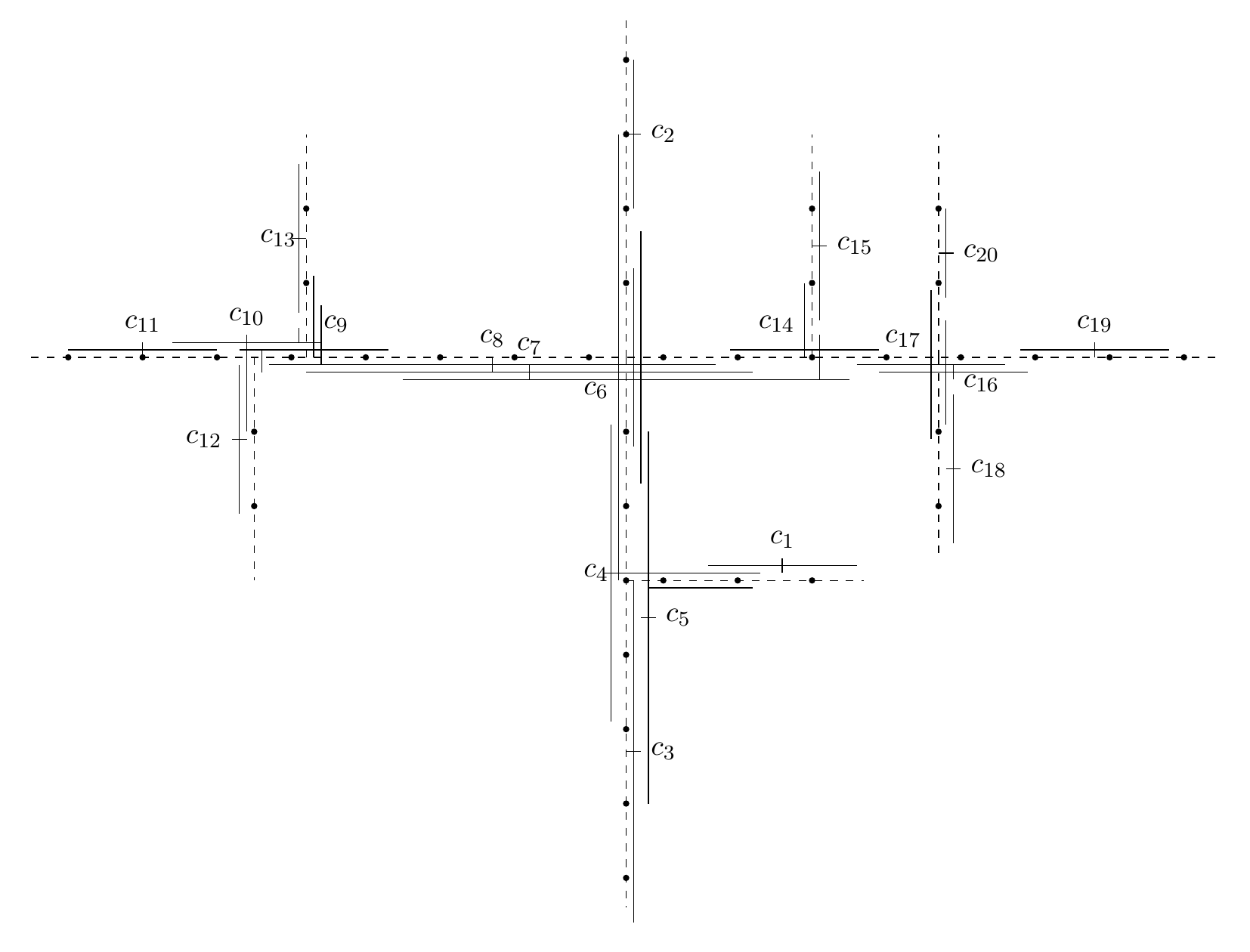}
\caption{In this figure, we illustrate an NSI model for the graph shown in Figure \ref{fig:stronglychordal}.}
\label{fig:NISModel}
\end{figure}

 Figure \ref{fig:stronglychordal} shows an NSI graph while Figure \ref{fig:NISModel} shows an NSI model for the same graph.
 Now, with all these definitions in place, we state our results. 

\begin{theorem}\label{thm:4statements}
The following three statements are equivalent for a graph $G$:
\begin{itemize}
\item[\textit{i)}] $G$ is NSI,
\item[\textit{ii)}] $G$ is proper-NSI,
\item[\textit{iii)}] $G$ is unit-NSI.
\end{itemize}
 \end{theorem}

 \begin{theorem}\label{thm:complete}
A complete signed graph  $S=(V,E^+\cup  E^-)$ has a valid distance drawing in a real tree if and only if its positive subgraph $S^+=(V,E^+)$ is NSI. 
 \end{theorem}
 
  
 \section{Related Work}\label{sec:relwork}
Kermarrec and Thraves Caro first introduced the SCFE problem in \cite{10.1007/978-3-642-22993-0_36}. They studied the SCFE problem in different metric spaces, such as $\mathbb{R}$ and $\mathbb{R}^2$ with the Euclidean distance. They exhibited signed graphs without valid distance drawings in $\mathbb{R}$ and $\mathbb{R}^2$. They also characterized the set of signed graphs with valid distance drawings in $\mathbb{R}$. Cygan et al. in \cite{cygan2015sitting} showed that the SCFE problem is NP-Complete when the metric space in consideration is $\mathbb{R}$ with the Euclidean distance. They also proved that a complete signed graph has a valid distance drawing in $\mathbb{R}$ if and only if its positive subgraph is a proper interval graph. 

After that, Pardo et al. in \cite{Pardo2015} studied an optimization version of the SCFE problem in the real line. The problem studied in \cite{Pardo2015} was to find an injection $D:V\rightarrow \mathbb{R}$ of the vertex set of a signed graph $S=(V,E^+\cup E^-)$ in $\mathbb{R}$ with the goal of minimizing the number of pairs of incident edges with different sign, $\{u,v\} \in E^+$, $\{u,w\}\in E^-$, such that $|D(u)-D(v)| \leq |D(u)-D(w)|$. They presented two heuristics using greedy techniques. They also showed a relationship between their optimization problem and the well-known Quadratic Assignment problem. Additional improvements on this optimization version of the problem were presented lately by Pardo et al. in \cite{pardo2020basic}.
 
Benitez et al. in \cite{benitez2018sitting} studied the SCFE problem in the circumference. They proved that it is NP-Hard to decide whether a given signed graph has a valid distance drawing in the circumference. Nevertheless, if the given signed graph is complete, they showed that it has a valid distance drawing if and only if its positive subgraph is a proper circular-arc graph, i.~e., the intersection graph of a family of arcs in the circumference where no arc is fully contained in another. 

Spaen et al. in \cite{Spaen2019} studied the SCFE problem from a different perspective. They studied the problem of finding $L(n)$, defined as the smallest dimension $k$ such that any signed graph on $n$ vertices has a valid distance drawing in $\mathbb{R}^k$ with the Euclidean distance. They showed that $\lfloor\log_5(n-3)\rfloor+1 \leq L(n) \leq n - 2$.

Aracena and Thraves Caro in \cite{aracena2019sitting} studied the SCFE problem for weighted graphs in the real line. In that case, a valid distance drawing is an injection that, for every pair of incident edges in the graph, the end vertices of the heavier edge are injected closer than the end vertices of the lighter edge. Given a weighted graph $G$, they constructed a polyhedron defined by a set of inequalities $M(G)\mathbf{x} \leq \mathbf{b}$. They proved that a  weighted graph $G$ has a valid distance drawing in $\mathbb{R}$ if and only if that polyhedron is not empty. The SCFE problem for weighted graphs in the real line is similar to the Seriation problem (see \cite{liiv2010seriation}). However, Aracena and Thraves Caro showed that the SCFE problem for weighted graphs in the real line and the Seriation problem are different. They showed that seriation is a necessary condition to solve the SCFE problem, but it is not sufficient. 

A graph is \emph{chordal} if all its cycles of four or more vertices have a \emph{chord}, which is an edge that is not part of the cycle but connects two vertices of the cycle. Gavril in \cite{GAVRIL197447} proved that chordal graphs are exactly the intersection graphs of subtrees in a tree. Therefore, NSI graphs are chordal. Tamir in \cite{tamir1983class} studied NSI graphs in the context of location problems. Let $G$ be an NSI graph, $A(G)$ be its adjacency matrix, and $A^*(G):=A(G) +I$ be its \emph{augmented adjacency} matrix, where $I$ denotes the identity. Arie Tamir in  \cite{tamir1983class} proved that for all NSI graph $G$, the matrix $A^*(G)$ is \emph{balanced}, where a balanced matrix is a $(0,1)$-matrix which does not contain any square submatrix of odd order having all row sums and all column sums equal to $2$. 

A direct consequence of Tamir's result is that an NSI graph is \emph{strongly chordal}, i. e. a chordal graph where every cycle of even length ($\geq 6$) has an edge that connects two vertices that are an odd distance apart from each other in the cycle. Nevertheless, not all strongly chordal graphs are NSI. Bibelnieks and Dearing in \cite{bibelnieks1993neighborhood} exhibited and example of a chordal graph that is not NSI. A \emph{Neighborhood Subtree Tolerance} (NeST) graph is a graph $G=(V,E)$ for which there exists a set of balls $\{T_i\}_{i\in V}$ and a set of tolerances $\{\tau_i\}_{i \in V}$ such that $\{i,j\} \in E \iff |T_i\cap T_j| \geq \min \{\tau_i, \tau_j\}$. A graph is \emph{constant} NeST if all the tolerances $\tau_i$ are equal to a constant $c$.  Bibelnieks and Dearing in \cite{bibelnieks1993neighborhood} proved that NSI graphs are exactly constant NeST graphs. 

 \section{NSI graphs, Proper-NSI graphs and Unit-NSI graphs are Equivalent}\label{sec:4statements}
In this section, we prove Theorem \ref{thm:4statements}. We recall the statement of the Theorem. 
\setcounter{theorem}{2}
\begin{theorem}\label{lem:3-statements}
The following three statements are equivalent for a graph $G$:
\begin{itemize}
\item[\textit{i)}] $G$ is NSI,
\item[\textit{ii)}] $G$ is proper-NSI,
\item[\textit{iii)}] $G$ is unit-NSI.
\end{itemize}
\end{theorem}
\begin{proof}
Let $G=(V,E)$ be a graph. We first show that \textit{iii)} implies \textit{ii)}. Assume that $G$ is a unit-NSI graph. Let $\mathcal{B}$ be $G$'s unit-NSI model. Since all balls in $\mathcal{B}$ have the same radius, $\mathcal{B}$ will also be a proper-NSI model unless there are two or more balls that are the same. Assume then that $\mathcal{B}$ has two or more balls that are the same. We show in the next two paragraphs that we can slightly modify the center of those repeated balls so that the intersection model maintains its intersection pattern and no ball is fully contained in another. Therefore, after this modification, we obtain a proper-NSI model. Hence, $G$ is a proper-NSI graph.

If two balls in $\mathcal{B}$ with positive radii intersect in only one point, they intersect in an extreme point of each other. Two balls that intersect only at an extreme point of each other are called two \emph{kissing} balls. If $\mathcal{B}$ has no kissing balls but has two or more balls that are the same, we proceed as follows. Let $B_1, B_2, \ldots B_k$ be the set of balls repeated in $\mathcal{B}$, i.~e., they have the same centers and radii. Let $\epsilon$ be the size of the smallest intersection between $B_1$ and any other ball in $\mathcal{B}$. On the other hand, let $\delta$ be the smallest distance between an extreme point of $B_1$ and any other extreme point of a ball in $\mathcal{B}$ which does not intersect with $B_1$. If we move the center of $B_1$ a distance strictly smaller than $\epsilon$ and $\delta$ in any direction, we will not create any new intersection between $B_1$ and another ball, and we will not remove any intersection between $B_1$ and another ball. Therefore, after this slightly modification of $B_1$, $\mathcal{B}$ maintains its intersection pattern and $B_1$ is not the same as $B_2, \ldots B_k$ anymore. Repeating this process, we obtain a proper-NSI model. 

If $\mathcal{B}$ has kissing balls, it has one ball that is kissing only one other ball. To see this, pick any ball that is kissing another ball. If it is kissing more than one ball, choose one of them and move to that ball. Repeat the process without returning to one ball that we have picked before. Due to the tree structure of $\mathcal{T}$, and since $\mathcal{B}$ is finite, this process will end in a ball that is kissing only one other ball. Let $B$ and $C$ be the two kissing balls, and $B$ be the one that is kissing only $C$. Now, let $\epsilon$ and $\delta$ be defined as in the previous paragraph for $B$, with the particularity that $\epsilon$ is now considered only between the intersections with positive size between $B$ and other balls. Move the center of $B$ towards the center of $C$, a distance strictly smaller than $\epsilon$ and $\delta$. Now $B$ and $C$ are not kissing balls, and the intersection pattern was not affected. Repeating this process, we obtain an intersection model with no kissing balls and with the same intersection pattern than $\mathcal{B}$.

If $G$ is a proper-NSI graph, it is also NSI. Therefore, \textit{ii)} implies \textit{i)}. Hence, we just need to show that \textit{i)} implies \textit{iii)}. 

Assume that $G$ is NSI, and let $\mathcal{T}$ be the real tree in which $G$ has its NSI model $\mathcal{B}$. Let $B(c_v,r_v)$ be the ball in $\mathcal{B}$ corresponding to vertex $v\in V$. Let $v^*$  be the vertex whose corresponding ball has the largest radius, denoted $r_{v^*}$. Now, for each ball $B(c_v,r_v)$ with radius strictly smaller than $r_{v^*}$, we add a branch (curve) to $\mathcal{T}$ at the point $c_v$. Then, we transform $B(c_v,r_v)$ by moving its center a distance $r_{v^*}-r_v$ from $c_v$ along the new added branch and increasing its radius to $r_{v^*}$.
Figure \ref{fig:NSItoUnitNSI} shows an example of this transformation.

\begin{figure}[t]
\begin{minipage}[t]{.33\linewidth}
 \centering\includegraphics[scale=0.5]{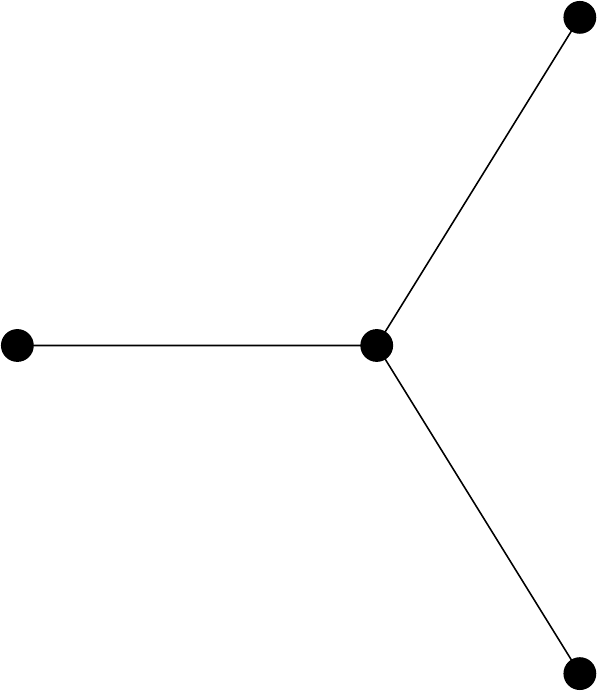}
 \subcaption{$K_{1,3}$}\label{fig:2:1}
 \end{minipage}
 \begin{minipage}[t]{.32\linewidth}
 \centering\includegraphics[scale=0.7]{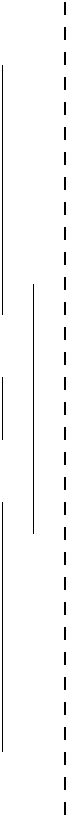}
 \subcaption{NSI model for $K_{1,3}$}\label{fig:2:2}
 \end{minipage}
 \begin{minipage}[t]{.32\linewidth}
 \centering\includegraphics[scale=0.7]{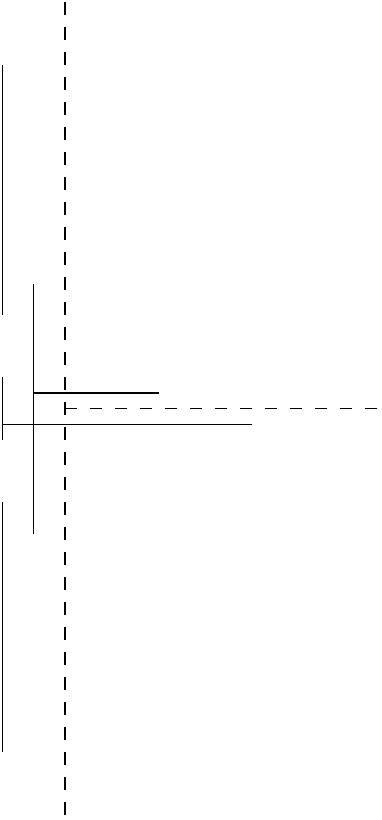}
 \subcaption{Unit-NSI model for $K_{1,3}$}\label{fig:2:3}
 \end{minipage}
 \caption{Subfigure \ref{fig:2:1} shows the $K_{1,3}$ bipartite grapha. Subfigure \ref{fig:2:2} shows an NSI model for $K_{1,3}$. Subfigure \ref{fig:2:3} shows a unit-NSI model for $K_{1,3}$. Figures \ref{fig:2:2} and \ref{fig:2:3} exemplify the transformation to go from an NSI model to a unit-NSI model described in the proof of Theorem \ref{thm:4statements}.
 \label{fig:NSItoUnitNSI}}
 \end{figure}

We do this process one ball at a time. We shall see that after each step, the intersection pattern does not change. Let $B(c_v,r_v)$ be the ball for which we modified the center and radius. Let $\mathcal{T}'$ be the new tree with the newly added branch ($\mathcal{T}$ is fully contained in $\mathcal{T}'$). Let $B'$ be the new ball in $\mathcal{T}'$ for $v$. It is worth noting that if a ball $C$ does not contain $c_v$ in $\mathcal{T}$, then $C$ is not modified in $\mathcal{T}'$. Therefore, since balls that do not intersect $B(c_v,r_v)$ do not contain $c_v$, they are not modified in $\mathcal{T}'$. It is worth noting that, balls that contain $c_v$ in $\mathcal{T}$ intersect with  $B(c_v,r_v)$.

The distance between the center of $B'$ and $c_v$ (the center of $B(c_v,r_v)$) is $r_{v^*}-r_v$. Since the radius of $B'$ is $r_{v^*}$, and the new branch is added in $c_v$, the intersection between $B'$ and $\mathcal{T}$ is equal to $B(c_v,r_v)$. Therefore, a ball in $\mathcal{T}$ intersects with  $B(c_v,r_v)$ if and only if it intersects with $B'$ in $\mathcal{T}'$. Consequently, the intersection pattern does not change after one step of this process. Repeating this modification for all balls with radius strictly smaller than $r_{v^*}$ we obtain an intersection model where all balls have the same radius, and that is still an intersection model for $G$. Hence, $G$ is a unit-NSI graph.
\end{proof}

 \section{Valid Distance Drawing in Trees for Complete Signed 
 Graphs}\label{sec:complete}
 In this section, we prove Theorem \ref{thm:complete}, which says that 
 a complete signed graph has a valid distance drawing in a tree if and 
 only if its positive subgraph is NSI. 
 We split the proof of this theorem in two lemmas. 

 \begin{lemma}
 Let $S$ be a complete signed graph. If the  
 positive subgraph $S^+$ of $S$ is NSI, 
 $S$ has a valid distance drawing in a tree. 
 \end{lemma}
 \begin{proof}
  Let $S=(V,E^+\cup E^-)$ be a complete signed graph such that the 
  positive subgraph $S^+=(V,E^+)$ of $S$ is NSI. 
  Using Theorem \ref{thm:4statements}, we have that
  $S^+$ is unit-NSI. Let 
  $\mathcal{B}=\{B(c_v,1):v\in V\}$ be $S^+$'s unit balls intersection 
  model in some tree 
  $(\mathcal{T},d)$. Consider  the injection $D:V\rightarrow 
  \mathcal{T}$ defined as follows: $D(v):=c_v$. 
  We claim that $D$ is a valid distance drawing for $S$ in 
  $\mathcal{T}$.
  To prove this claim, consider two vertices $u$ and $v$ in $V$. 
  Note that $d(c_u,c_v) \leq 2$
  if $u$ and $v$ 
  are neighbors, since their balls intersect. Otherwise, $d(c_u,c_v) > 
  2$. Therefore, 
  if $\{u,v\}\in E^+$ and $\{u,w\}\in E^-$ are two incident edges with 
  different signs, we have that 
  $d(c_u,c_v) \leq 2 < d(c_u,c_w)$. Hence, $D$ is a valid distance 
  drawing for $S$ in $\mathcal{T}$.
\end{proof}

 \begin{lemma}\label{lem:valid-to-strngly}
 Let $S$ be a complete signed graph. If $S$ has a valid 
 distance drawing in a tree, 
 the positive subgraph $S^+$ of $S$ is NSI.  
 \end{lemma}
 \begin{proof}
 Let $S=(V,E^+\cup E^-)$ be a complete signed graph with a valid 
 distance drawing in a tree $(\mathcal{T},d)$. 
 Let $D:V\rightarrow \mathcal{T}$ be a valid distance drawing for $S$ in
 $\mathcal{T}$.
 For each vertex $v \in V$, we use $v^+$ to denote its farthest friend, i.~e., $v^+$ is the vertex $u \in N^+[v]$ that maximizes $d(v,u)$. 
 %
 We define the following family of balls in $\mathcal{T}$, 
 \[\mathcal{B}:=\bigg\{B\left(D(v),\frac{d(D(v),D(v^+))}{2}\right):v\in V\bigg\}.
\]
We claim that the intersection graph of $\mathcal{B}$ is equal to $S^+$. 
 To prove this claim, we show that two balls in $\mathcal{B}$ intersect if and only if 
 their corresponding vertices are friends. 
 
 Let $u$ and $v$ be two vertices in $V$ such that: \[B\left(D(v),\frac{d(D(v),D(v^+))}{2}\right) \bigcap B\left(D(u),\frac{d(D(u),D(u^+))}{2}\right)\neq \emptyset.\] 
 Therefore, 
 \begin{eqnarray*}
 d(D(u),D(v)) &\leq& \frac{d(D(u),D(u^+))}{2} + \frac{d(D(v),D(v^+))}{2}\\ 
 &\leq& \max\{d(D(u),D(u^+)),d(D(v),D(v^+)) \}.
 \end{eqnarray*}
Hence, either $u$ is closer to $v$ than its farthest friend or $v$ is closer to $u$ than its farthest friend. 
In any of these two cases, since $D$ is a valid distance drawing, we can conclude than $u$ and $v$ are friends. 

Now, assume that $u$ and $v$ are friends. Then, 
\begin{eqnarray*}
d(D(u),D(v)) &\leq& \min\{d(D(u),D(u^+)),d(D(v),D(v^+)) \}\\
&\leq& \frac{d(D(u),D(u^+))}{2} + \frac{d(D(v),D(v^+))}{2},
 \end{eqnarray*}
 which allows us to conclude that
 \[B\left(D(v),\frac{d(D(v),D(v^+))}{2}\right) \bigcap B\left(D(u),\frac{d(D(u),D(u^+))}{2}\right)\neq \emptyset.\] 
 Therefore, $S^+$ is NSI. 
\end{proof} 
 
\begin{figure}[t]
\centering
\includegraphics[width=\textwidth]{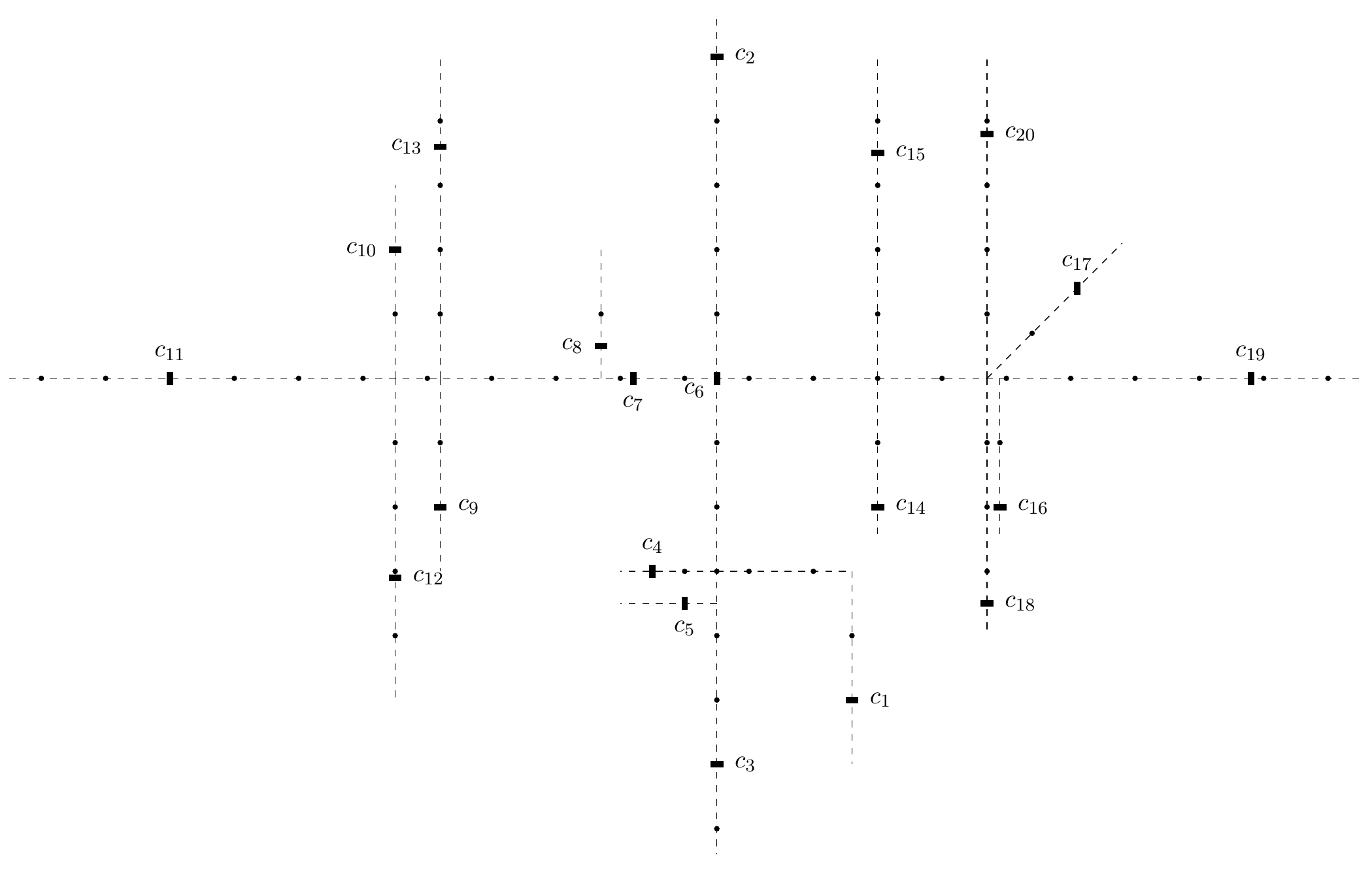}
\caption{In this figure, we illustrate a valid distance drawing in a tree for the complete signed graph whose positive subgraph is the graph shown in Figure \ref{fig:stronglychordal}.}
\label{fig:vdd}
\end{figure}

Now, if we put together these two lemmas we obtain Theorem \ref{thm:complete}.
Figure \ref{fig:vdd} shows a valid distance drawing in a tree for the complete signed graph whose positive subgraph is the graph of the example in Figure \ref{fig:stronglychordal}.

 \section{Concluding Remarks}\label{sec:conclusion}
We conclude this paper by presenting two interesting open problems. The first problem is to find a combinatorial characterization for NSI graphs. In the same line, are NSI graphs recognizable in polynomial time?.  

The second question is related to the SCFE problem in trees. Is it NP-Complete to decide the existence of a valid distance drawing in a tree when the signed graph is not necessarily complete?. Previous results show some evidence in that line. Indeed, if we know that a signed graph $G$, not necessarily complete, has a valid distance drawing in a tree, finding a tree $T$ with the minimum number of leaves such that $G$ has a valid distance drawing in $T$ is an NP-Hard problem due to the complexity result presented in \cite{cygan2015sitting}.

\end{document}